# Application of Data Mining In Marketing

[1] Radhakrishnan B, [2] Shineraj G, [3] Anver Muhammed K.M

[1, 2, 3] Dept. of Computer Science, Baselios Mathews II College of Engineering, Kerala, India

**Abstract**
One of the most important problems in modern finance is finding efficient ways to summarize and visualize the stock market data to give individuals or institutions useful information about the market behavior for investment decisions. The enormous amount of valuable data generated by the stock market has attracted researchers to explore this problem domain using different methodologies. Potential significant benefits of solving these problems motivated extensive research for years. The research in data mining has gained a high attraction due to the importance of its applications and the increasing generation information. This paper provides an overview of application of data mining techniques such as decision tree. Also, this paper reveals progressive applications in addition to existing gap and less considered area and determines the future works for researchers.

## 1. Introduction

Data mining, as we use the term, is the exploration and analysis of large quantities of data in order to discover meaningful patterns and rules. the *goal* of data mining is to allow a corporation to improve its marketing, sales, and customer support operations through a better understanding of its customers. Data mining comes in two flavors directed and undirected. Directed data mining attempts to explain or categorize some particular target field such as income or response. Undirected data mining attempts to find patterns or similarities among groups of records without the use of a particular target field or collection of predefined classes.

Many problems of intellectual, economic, and business interest can be phrased in terms of the following six tasks:

- ■ Classification
- ■ Estimation
- ■ Prediction
- ■ Affinity grouping
- ■ Clustering
- ■ Description and profiling

### 1.1 Classification

Classification consists of examining the features of a newly presented object and assigning it to one of a predefined set of classes. The objects to be classified are generally represented by records in a database table or a file, and the act of classification consists of adding a new column with a class code of some kind.

### 1.2 Estimation

Estimation deals with continuously valued outcomes. Given some input data, estimation comes up with a value for some unknown continuous variable such as income, height, or credit card balance. estimation is often used to perform a classification task.

### 1.3 Prediction

Prediction is the same as classification or estimation, except that the records are classified according to some predicted future behavior or estimated future value. In a prediction task, the only way to check the accuracy of the classification is to wait and see. The primary reason for treating prediction as a separate task from classification and estimation is that in predictive modeling there are additional issues regarding the temporal relationship of the input variables or predictors to the target variable.

### 1.4 Affinity Grouping

The task of affinity grouping is to determine which things go together. The prototypical example is determining what things go together in a shopping cart at the supermarket, the task at the heart of *market basket analysis*. Retail chains can use affinity grouping to plan the arrangement of items on store shelves or in a catalog so that items often purchased together will be seen together.

### 1.5 Clustering

Clustering is the task of segmenting a heterogeneous population into a number of more homogeneous subgroups or *clusters*. In clustering, there are no predefined classes and no examples. The records are grouped together on the basis of self-similarity. It is up to the user to determine what meaning, if any, to attach to the resulting clusters. Clusters of symptoms might indicate different diseases. K-means clustering is an exclusive clustering algorithm. Each object is assigned to exactly one of a set of clusters. For this method of clustering we start by deciding how many clusters we would like to form from our data. We call this value j. The value of j is generally a small integer, such as 1,2 or 3, or can be larger. There are many ways in which k clusters might potentially be formed. We can measure the quality of a set of clusters by taking the sum of the squares of the distances of each point from the centroid of the cluster to which it is assigned. We would like the value of this function to be as small as possible. We next select j .These are treated as the centroids of j clusters, or to be more precise as the centroids of j potential clusters, which at present have no members. We can select these points in any way we wish, but the method may work better if we select k initial points that are fairly far apart. We now assign each of the points one by one to the cluster which has the nearest centroid. When all the objects have been assigned we will have j clusters based on the original j centroids but the 'centroids' will no longer be the true centroids of the clusters. Next we recalculate the centroids of the clusters, and then repeat the previous steps, assigning each object to the cluster with the nearest centroid etc

1.6 Profiling

Sometimes the purpose of data mining is simply to describe what is going on in a complicated database in a way that increases our understanding of the people, products, or processes that produced the data in the first place. A good enough *description* of a behavior will often suggest an *explanation* for it as well. Decision trees are a powerful tool for profiling customers with respect to a particular target or outcome.

Data mining makes the most sense when there are large volumes of data. In fact, most data mining algorithms *require* large amounts of data in order to build and train the models that will then be used to perform classification, prediction, estimation, or other data mining tasks. Data warehousing brings together data from many different sources in a common format with consistent definitions for keys and fields. It is generally not possible (and certainly not advisable) to perform computer- and input/output (I/O)–intensive data mining operations on an operational system that the business depends on to survive.

Data mining is being used to promote customer retention in any industry where customers are free to change suppliers at little  cost and competitors are eager to lure them away. Banks call it attrition. Wireless phone companies call it churn. By any name, it is a big problem. By gaining an understanding of *who* is likely to leave and *why*, a retention plan can be developed that addresses the right issues and targets the right customers. In many industries, some customers cost more than they are worth. These might be people who consume a lot of customer support resources without buying much. Or, they might be those annoying folks who carry a credit card they rarely use, are sure to pay off the full balance when they do, but must still be mailed a statement every month.

## 2. Identify the Business Opportunity

The virtuous cycle of data mining starts with identifying the right business opportunities. Unfortunately, there are too many good statisticians and competent analysts whose work is essentially wasted because they are solving problems that don't help the business. Good data miners want to avoid this situation.

Avoiding wasted analytic effort starts with a willingness to act on the results. Many normal business processes are good candidates for data mining:

- Planning for a new product introduction
- Planning direct marketing campaigns
- Understanding customer attrition/churn
- Evaluating results of a marketing test

## 3. Mining Data

Data mining, transforms data into actionable results. Success is about making business sense of the data, not using particular algorithms or tools. Numerous pitfalls interfere with the ability to use the results of data mining:

- Bad data formats, such as not including the zip code in the customer address in the results.
- Confusing data fields, such as a delivery date that means "planned delivery date" in one system and "actual delivery date" in another system.

Data comes in many forms, in many formats, and from multiple systems. The virtuous cycle of data mining starts with identifying the right business opportunities. Identifying the right data sources and bringing them together are critical success factors. Every data mining project has data issues: inconsistent systems, table keys that don't match across databases, records overwritten every few months, and so on. Taking action is the purpose of the virtuous cycle of data mining. Data mining makes business decisions more informed. Over time, we expect that better-informed decisions lead to better results. measurements provide information for making more informed decisions in the future. Data mining is about connecting the past—through learning to future actions.

## 4. Using Data Mining in Marketing

The simplest definition of a good prospect—and the one used by many companies—is simply someone who might at least express interest in becoming a customer. Data mining is applied to this problem by first defining what it means to be a good prospect and then finding rules that allow people with those characteristics to be targeted. Prospecting requires communication. Broadly speaking, companies intentionally communicate with prospects in several ways. One way of targeting prospects is to look for people who resemble current customers. For instance, through surveys, one nationwide publication determined that its readers have the following characteristics:

- 59 percent of readers are college educated.
- 46 percent have professional or executive occupations.
- 21 percent have household income in excess of $75,000/year.
- 7 percent have household income in excess of $100,000/year.

One way of determining whether a customer fits a profile is to measure the similarity which we also call distance between the customer and the profile. Several data mining techniques use this idea of measuring similarity as a distance. Consider two survey participants. Amy is college educated, earns $80,000/year, and is a professional. Bob is a high-school graduate earning $50,000/year. Which one is a better match to the readership profile? The answer depends on how the comparison is made. Table1 shows one way to develop a score using only the profile and a simple distance metric. This table calculates a score based on the proportion of the audience that agrees with each characteristic. For instance, because 58 percent of the readership is college educated, Amy gets a score of 0.58 for this characteristic. Bob, who did not graduate from college, gets a score of 0.42 because the other 42 percent of the readership presumably did not graduate from college. This is continued for each characteristic, and the scores are added together. Amy ends with a score of 2.18 and Bob with the higher score of 2.68. His higher score reflects the fact that he is more similar to the profile of current readers than is Amy.

Table 1. Calculating Fitness Scores for Individuals by Comparing Them along Each Demographic Measure

|  | Readership | Yes Score | No Score | AMY | BOB |
|---|---|---|---|---|---|
| College Educated | 58% | 0.58 | 0.42 | 0.58 | 0.42 |
| Proof or exec | 46% | 0.46 | 0.54 | 0.46 | 0.54 |
| Income greater than $75K | 21% | 0.21 | 0.79 | 0.21 | 0.79 |
| Income greater than $100K | 0.07 | 0.93 | 0.93 | 0.93 | 0.93 |
| Total |  |  |  | 2.18 | 2.68 |

### 4.1 Data Mining to Improve Direct Marketing Campaigns

Advertising can be used to reach prospects about whom nothing is known as individuals. Direct marketing requires at least a tiny bit of additional information such as a name and address or a phone number or an email address. Where there is more information, there are also more opportunities for data mining. At the most basic level, data mining can be used to improve targeting by selecting which people to contact. the first level of targeting does not require data mining, only data. Direct marketing campaigns typically have response rates measured in the single digits. Response models are used to improve response rates by identifying prospects that are more likely to respond to a direct solicitation.

With existing customers, a major focus of customer relationship management is increasing customer profitability through cross-selling and up-selling. Data mining is used for figuring out what to offer to whom and when to offer it. Charles Schwab, the investment company, discovered that customers generally open accounts with a few thousand dollars even if they have considerably more stashed away in savings and investment accounts. Naturally, Schwab would like to attract some of those other balances. By analyzing historical data, they discovered that customers who transferred large balances into investment accounts usually did so during the first few months after they opened their first account. After a few months, there was little return on trying to get customers to move in large balances. The window was closed. As a result of learning this, Schwab shifted its strategy from sending a constant stream of solicitations throughout the customer life cycle to concentrated efforts during the first few months.

Customer attrition is an important issue for any company, and it is especially important in mature industries where the initial period of exponential growth has been left behind. Not surprisingly, churn (or, to look on the bright side, retention) is a major application of data mining. One of the first challenges in modeling churn is deciding what it is and recognizing when it has occurred. When a once loyal customer deserts his regular coffee bar for another down the block, the barista who knew the customer's order by heart may notice, but the fact will not be recorded in any corporate database. Even in cases where the customer is identified by name, it may be hard to tell the difference between a customer who has churned and one who just hasn't been around for a while. If a loyal Ford customer who buys a new F150 pickup every 5 years hasn't bought one for 6 years, can we conclude that he has defected to another brand. Churn is important because lost customers must be replaced by new customers, and new customers are expensive to acquire and generally generate less revenue in the near term than established customers. There are two basic approaches to modeling churn. The first treats churn as a binary outcome and predicts which customers will leave and which will stay. The second tries to estimate the customers' remaining lifetime.

### 4.2 Data Mining using familiar tools

The *null hypothesis* is the assumption that differences among observations are due simply to chance. The null hypothesis is not only an approach to analysis; it can also be quantified. The *p-value* is the probability that the null hypothesis is true. Remember, when the null hypothesis is true, nothing is really happening, because differences are due to chance. A *statistic* refers to a measure taken on a sample of data. Statistics is the study of these measures and the samples they are measured on. A good place to start, then, is with such useful measures, and how to look at data. The most basic descriptive statistic about discrete fields is the number of times different values occur. A histogram shows how often each value occurs in the data and can have either absolute quantities (204 times) or percentage (14.6 percent). Histograms are quite useful and easily made with Excel or any statistics package. However, histograms describe a single moment.

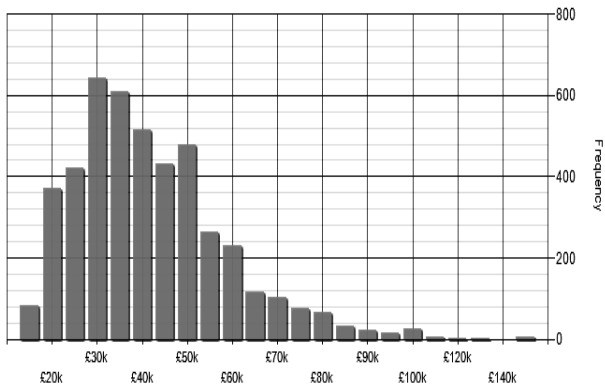

Figure 1: Histogram diagram

Data mining is often concerned with what is happening over time. Time series analysis requires choosing an appropriate time frame for the data; this includes not only the units of time, but also when we start counting from. Some different time frames are the beginning of a customer relationship, when a customer requests a stop, the actual stop date, and so on. The next step is to plot the time series. A time series chart provides useful information. However, it does not give an idea as to whether the changes over time are expected or unexpected. For this, we need some tools from statistics. One way of looking at a time series is as a partition of all the data, with a little bit on each day. There is a basic theorem in statistics, called the Central Limit Theorem "As more and more samples are taken from a population, the distribution of the averages of the samples (or a similar statistic) follows the normal distribution. The average (what statisticians call the mean) of the samples comes arbitrarily close to the average of the entire population".

The normal distribution is described by two parameters, the mean and the standard deviation. The mean is the average count for each day. The standard deviation is a measure of the extent to which values tend to cluster around the mean. Assuming that the standardized value follows the normal distribution makes it possible to calculate the probability that the value would have occurred by chance. Actually, the approach is to calculate the probability that something further from the mean would have occurred—the p-value. The reason the exact value is not worth asking is because any given z-value has an arbitrarily small probability. Probabilities are defined on ranges of z-values as the area under the normal curve between two points. Calculating something further from the mean might mean either of two things:

1. The probability of being more than *z* standard deviations from the mean.

2. The probability of being *z* standard deviations greater than the mean.

The first is called a two-tailed distribution and the second is called a one tailed distribution.

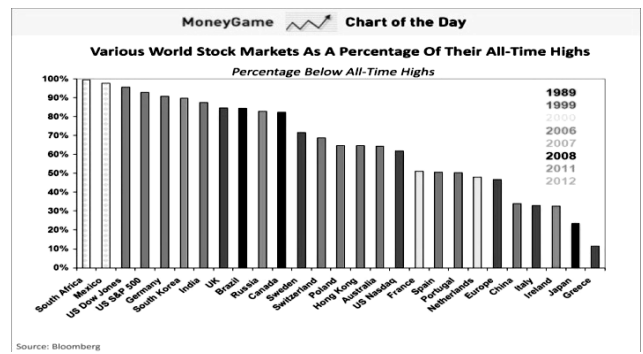

Figure2: A time chart can also be used for continuous values.

Time series are an example of cross-tabulation—looking at the values of two or more variables at one time. For time series, the second variable is the time something occurred. The most commonly used statistic is the *mean* or average value (the sum of all the values divided by the number of them).

Some other important things to look at are:

**Range.** The range is the difference between the smallest and largest observation in the sample. The range is often looked at along with the minimum and maximum values themselves.

**Mean.** This is what is called an average in everyday speech.

**Median.** The median value is the one which splits the observations into two equally sized groups, one having observations smaller than the median and another containing observations larger than the median.

**Mode.** This is the value that occurs most often.

Variance is a measure of the dispersion of a sample or how closely the observations cluster around the average. The range is not a good measure of dispersion because it takes only two values into account—the extremes. Removing one extreme can, sometimes, dramatically change the range. The variance, on the other hand, takes every value into account. The difference between a given observation and the mean of the sample is called its *deviation*. The variance is defined as the average of the squares of the deviations. Standard deviation, the square root of the variance, is the most frequently used measure of dispersion. It is more convenient than variance because it is expressed in the same units as the observations rather than in terms of those units squared. This allows the standard deviation itself to be used as a unit of measurement.

The z-score, is an observation's distance from the mean measured in standard deviations. Using the normal distribution, the z-score can be converted to a probability or confidence level. *Correlation* is a measure of the extent to which a change in one variable is related to a change in another. Correlation ranges from –1 to 1. A correlation of 0 means that the two variables are not related. A correlation of 1 means that as the first variable changes, the second is guaranteed to change in the same direction, though not necessarily by the same amount. Another measure of correlation is the R2 value, which is the correlation squared and goes from 0 (no relationship) to 1 (complete relationship).

4.3 Decision Tree

A decision tree is a structure that can be used to divide up a large collection of records into successively smaller sets of records by applying a sequence of simple decision rules. With each successive division, the members of the resulting sets become more and more similar to one another. A decision tree model consists of a set of rules for dividing a large heterogeneous population into smaller, more homogeneous groups with respect to a particular target variable.

Decision trees can also be used to estimate the value of a continuous variable, although there are other techniques more suitable to that task. A record enters the tree at the root node. The root node applies a test to determine which *child node* the record will encounter next. There are different algorithms for choosing the initial test, but the goal is always the same: To choose the test that best discriminates among the target classes. This process is repeated until the record arrives at a *leaf node*. All the records that end up at a given leaf of the tree are classified the same way. There is a unique path from the root to each leaf. That path is an expression of the *rule* used to classify the records. Different leaves may make the same classification, although each leaf makes that classification for a different reason.

The decision tree can be used to answer that question too. Assuming that order amount is one of the variables available in the pre classified model set, the average order size in each leaf can be used as the estimated order size for any unclassified record that meets the criteria for that leaf. It is even possible to use a numeric target variable to build the tree; such a tree is called a *regression tree*. Instead of increasing the purity of a categorical variable, each split in the tree is chosen to decrease the variance in the values of the target variable within each child node. Although there are many variations on the core decision tree algorithm, all of them share the same basic procedure: Repeatedly split the data into smaller and smaller groups in such a way that each new generation of nodes has greater purity than its ancestors with respect to the target variable. At the start of the process, there is a training set consisting of pre classified records—that is, the value of the target variable is known for all cases. The goal is to build a tree that assigns a class (or a likelihood of membership in each class) to the target field of a new record based on the values of the input variables.

The tree is built by splitting the records at each node according to a function of a single input field. The first task, therefore, is to decide which of the input fields makes the best split. The best split is defined as one that does the best job of separating the records into groups where a single class predominates in each group. The measure used to evaluate a potential split is *purity*.

## 5. Conclusion

With the increase of economic globalization and evolution of information technology, financial data are being generated and accumulated at an unprecedented pace. As a result, there has been a critical need for automated approaches to effective and efficient utilization of massive amount of financial data to support companies and individuals in strategic planning and investment decision making. Data mining techniques have been used to uncover hidden patterns and predict future trends and behaviors in financial markets. The competitive advantages achieved by data mining include increased revenue, reduced cost, and much improved marketplace responsiveness and awareness. This paper therefore recommends various organizations to use data mining techniques in future to resolve complex problems.